\documentclass[aps,prl,preprint]{revtex4}

\usepackage{epsfig}
\draft

\begin{document}
\author{Feng-yao Hou$^{1}$, Lei Chang$^{2}$, Wei-min Sun$^{3,1}$, Hong-shi Zong$^{4,1,3}$, Yu-xin Liu$^{2,4}$}
\address{$^{1}$ Department of Physics, Nanjing University, Nanjing 210093, China}
\address{$^{2}$ Department of physics, Peking University, Beijing 100871, China}
\address{$^{3}$ Joint Center for Particle, Nuclear Physics and Cosmology, Nanjing 210093, China}
\address{$^{4}$ CCAST(World Laboratory), P.O. Box 8730, Beijing 100080, China}

\title{New Method for Numerically Solving the Chemical Potential Dependence of the Dressed Quark Propagator}

\begin{abstract}
Based on the rainbow approximation of Dyson-Schwinger equation and
the assumption that the inverse dressed quark propagator at finite chemical potential is analytic in the neighborhood of $\mu=0$, a new method for obtaining the dressed quark propagator at finite chemical potential $\mu$ from the one at zero chemical potential is developed. Using this method the dressed quark propagator at finite chemical potential can be obtained directly from the one at zero chemical potential without the necessity of
numerically solving the corresponding coupled integral equations by iteration methods. A comparison with previous results is given.

\bigskip

Key-words: Chemical potential dependence, Dyson-Schwinger approach.

\bigskip

E-mail: zonghs@chenwang.nju.edu.cn.

\bigskip

PACS Numbers: 24.85.+p; 11.10.Wx;12.39.Ba;14.20.Dh

\end{abstract}

\maketitle

QCD at finite density is of fundamental importance, both on purely theoretical and phenomenological grounds. Although considerable progress has been achieved in the lattice treatment of finite density QCD[1-3] recently, analytical studies with effective theories are useful alternatives and often shed more light on nonperturbative phenomena than the lattice approach. Due to the fact that the quark propagator at finite chemical potential plays an essential role in the study of chiral symmetry restoration and quark deconfinement, it is interesting to give a general recipe to study the chemical potential dependence of the dressed-quark propagator
at non-zero chemical potential in the framework of a suitable
nonperturbative QCD model.

Since the Dyson-Schwinger(DS) approaches[4-6] provides a
nonperturbative framework which admits the simultaneous study of
dynamical chiral symmetry breaking and confinement, it is
expected to be well suited to explore the transition from hadronic
matter to QGP[7]. It is the aim of this letter to study the
chemical potential dependence of the dressed-quark propagator in
the framework of DS approach, which provides a means of determining
the behavior of the chiral and deconfinement order parameters. Up to this end let us  briefly review the DS approach used in this letter. The DS approach is based on a coupled set of integral equations between the quark, gluon, and ghost propagators and vertex functions. They form a countably infinite set of coupled integral equations with the equation for an n-point Schwinger function depending on (n+1) and higher point functions.
In order to handle this system it is necessary to make certain simplifications and truncations. One approach, which is commonly referred to as the ``rainbow'' approximation, employs the bare quark-gluon vertex and solves the Dyson-Schwinger equation(DSE) for the dressed quark propagator, with a given dressed gluon propagator $g^2_{s}D_{\mu\nu}(p)$ as input. For example, under the ``rainbow'' approximation, the quark self-energy $\Sigma[\mu](p)$ at the non-zero chemical potential $\mu$ can be written as
\begin{equation}
\Sigma[\mu](p)=\frac{4}{3}\int \frac{d^4 q}{(2\pi)^4} g^2_{s}D_{\mu\nu}(p-q)\gamma_{\mu}{\cal{G}}[\mu](q)\gamma_{\nu}.
\end{equation}
The quark propagator ${\cal{G}}[\mu](p)$ and the quark self-energy $\Sigma[\mu](p)$ are related by
\begin{eqnarray}
{\cal{G}}^{-1}[\mu](p)&=&i\gamma\cdot p-\mu\gamma_4+\Sigma[\mu](p)\nonumber\\
&=&i\gamma\cdot p-\mu\gamma_4+\frac{4}{3}\int \frac{d^4 q}{(2\pi)^4} g^2_{s}D_{\mu\nu}(p-q)\gamma_{\mu}{\cal{G}}[\mu](q)\gamma_{\nu}.
\end{eqnarray}
Here we want to stress that Eq.(2), which employs the bare quark-gluon vertex and solves the DSE for the dressed quark propagator ${\cal{G}}[\mu](p)$ at finite chemical potential with a given chemical potential independent gluon propagator $g^2_{s}D_{\mu\nu}(p)$ as input, is our starting point for studying the dressed quark propagator at finite chemical potential in the framework of DS approach. 

If $\mu$ is set to be zero, ${\cal{G}}[\mu](p)$ goes into the dressed quark propagator in the zero chemical potential $G(p)\equiv{\cal{G}}[\mu=0](p)$, which reads
\begin{equation}
G^{-1}(p)\equiv i\gamma\cdot pA(p^2)+B(p^2)=i\gamma\cdot p+\frac{4}{3}\int \frac{d^4 q}{(2\pi)^4} g^2_{s}D_{\mu\nu}(p-q)\gamma_{\mu}G(q)\gamma_{\nu},
\end{equation}
with the self energy functions $A(p^2)$ and $B(p^2)$ being
determined by the rainbow DSE in the chiral limit:
\[
[A(p^2)-1]p^2=\frac{8}{3}\int \frac{d^{4}q}{(2\pi)^4}g^2_{s} D(p-q)
\frac{A(q^2)p\cdot q}{q^2A^2(q^2)+B^2(q^2)},
\]
\begin{equation}
B(p^2)=\frac{16}{3}\int \frac{d^{4}q}{(2\pi)^4}g^2_{s} D(p-q)
\frac{B(q^2)}{q^2A^2(q^2)+B^2(q^2)},
\end{equation}
for convenience, we have used a model ansatz 
$g^2_{s} D_{\mu\nu}(p)=\delta_{\mu\nu}g^2_{s} D(p^2)$ for the gluon propagator, which is often referred to as the so-called ``Feynman-like'' gauge propagator[5]. It is readily seen that the $B(p^2)$ in Eq.(4) has two qualitatively distinct solutions. The ``Nambu-Goldstone'' solution, for which $B(p^2)\neq 0$,
describes a phase, in which: 1) chiral symmetry is dynamically
broken, because one has a nonzero quark mass function; and 2) the
dressed quarks are confined, because the propagator described by
these functions does not have a Lehmann representation. The
other solution, the ``Wigner'' one, $B(p^2)\equiv 0$,
describes a phase, in which chiral symmetry is not broken and the dressed-quarks are not confined[4,5]. In the ``Wigner'' phase, the Dyson-Schwinger equation(4) reduces to:
\begin{equation}
[A'(p^2)-1]p^2=\frac{8}{3}\int \frac{d^{4}q}{(2\pi)^4}g_{s}^2 D(p-q)
\frac{p\cdot q}{q^2A'(q^2)},
\end{equation}
where $A'(p^2)$ denotes the dressed quark vector self energy function in the ``Wigner'' phase.  

In order to motivate our new method for calculating the dressed quark propagator at finite chemical potential, let us first recall the methods adopted in the previous literatures[8-11]. From Lorentz structure analysis, the most general form for the dressed quark propagator at finite chemical potential reads
\begin{equation}
{\cal{G}}^{-1}[\mu](p)=i\vec{\gamma}\cdot\vec{p}{\cal{A}}(\tilde{p})+i\gamma_4{\cal{C}}(\tilde{p})+{\cal{B}}(\tilde{p})-\mu\gamma_4\vec{\gamma}\cdot\vec{p}{\cal{D}}(\tilde{p}),
\end{equation} 
where ${\tilde p}=({\vec p}, p_4+i\mu)$. Substituting Eq.(6) into Eq.(2) and by means of suitable projection procedure(multiplying by appropriate gamma matrices and then taking the trace), the four independent scalar function ${\cal{A}}(\tilde{p})$, ${\cal{B}}(\tilde{p})$, ${\cal{C}}(\tilde{p})$, and ${\cal{D}}(\tilde{p})$ are found to satisfy a coupled set of Dyson-Schwinger equations. 

In principle, for a given model gluon propagator, one can numerically solve these coupled DSEs. However, the numerical solution for an arbitrary gluon propagator turns out to be rather difficult in practice(In Ref.[10], the last term on the right hand of Eq.(6) was dropped). In order to avoid the difficulty of numerically solving four independent coupled integral equations, we have proposed a new method in Ref.[12] for calculating the chemical potential dependence of the dressed quark propagator. For the sake of completeness we outline the main points of this method here.  Let us assume that the full inverse quark propagator at finite chemical potential is analytic in $\mu$, at least 
for small $\mu$. Under this assumption, one can expand ${\cal{G}}^{-1}[\mu](p)$ in powers of $\mu$ as follows[12]
\begin{eqnarray}
{\cal{G}}^{-1}[\mu]&&=\left.{\cal{G}}^{-1}[\mu]\right|_{\mu=0}+\left.
\frac{\partial {\cal{G}}^{-1}[\mu]}{\partial {\mu}} \right|_{\mu=0}\mu+
\cdots+\left.\frac{1}{n!}\frac{\partial^n {\cal{G}}^{-1}[\mu]}{\partial {\mu^n}} \right|_{\mu=0}\mu^n+\cdots\nonumber\\
&&=G^{-1}+\Gamma^{(1)}\mu+\cdots+\frac{1}{n!}\Gamma^{(n)}\mu^n+\cdots,
\end{eqnarray}
with $\Gamma^{(n)}(p,0)$
\begin{equation}
\Gamma^{(n)}(p,0)=\left. \frac{\partial^n {\cal{G}}^{-1}[\mu](p)}{\partial \mu^n} \right|_{\mu=0}~~~~n \geq 1.
\end{equation}
It should be noted that Eq.(7) is only valid within the radius of convergence of $\mu$ expansion. 

Working in the framework of rainbow approximation of DSE, one can prove the following by induction[12]
\begin{equation}
\Gamma^{(n)}(p,0)\equiv\frac{\partial^n G^{-1}(p)}{\partial(-ip_4)^n}, ~~~~n \geq 1.
\end{equation}

Combining Eqs.(7) and (9) we have
\begin{eqnarray}
{\cal G}^{-1}[\mu](p)&=&G^{-1}(p)+\Gamma^{(1)}(p,0)\mu+\cdots+\frac{1}{n!}\Gamma^{(n)}(p,0)\mu^n+\cdots \nonumber \\
&=& G^{-1}(p)+\frac{\partial G^{-1}(p)}{\partial(-ip_4)}\mu+\cdots+\frac{1}{n!}\frac{\partial^n G^{-1}(p)}{\partial(-ip_4)^n}\mu^n+\cdots  \nonumber \\
&=& G^{-1}(p)+\frac{\partial G^{-1}(p)}{\partial(p_4)}i\mu+\cdots+\frac{1}{n!}\frac{\partial^n G^{-1}(p)}{\partial(p_4)^n}(i\mu)^n+\cdots \nonumber \\
&=& G^{-1}({\vec p}, p_4+i\mu)\equiv G^{-1}({\tilde p})=i\gamma\cdot\tilde{p}A(\tilde{p}^2)+B(\tilde{p}^2).
\end{eqnarray}
Therefore the dressed quark propagator at finite $\mu$ can be obtained from that at zero $\mu$ by the simple substitution $p_4 \rightarrow p_4+i\mu$. From this result one sees
immediately that there are only two independent Lorentz structures(instead of four, which is determined from general analysis) in the dressed quark propagator at finite chemical potential.  

Substituting Eq.(10) into Eq.(2) and then multiplying by appropriate gamma matrices and taking the trace in Eq.(2), we have the following:
\[
[A(\tilde{p}^2)-1]\tilde{p}^2=\frac{8}{3}\int \frac{d^{4}q}{(2\pi)^4}g^2_{s} D(p-q)
\frac{A(\tilde{q}^2)\tilde{p}\cdot\tilde{q}}{q^2A^2(\tilde{q}^2)+B^2(\tilde{q}^2)},
\]
\begin{equation}
B(\tilde{p}^2)=\frac{16}{3}\int \frac{d^{4}q}{(2\pi)^4}g^2_{s} D(p-q)
\frac{B(\tilde{q}^2)}{q^2A^2(\tilde{q}^2)+B^2(\tilde{q}^2)}.
\end{equation}
In principle, for a given model gluon propagator, one can numerically solve this set of coupled DSEs using iteration methods just as one solves the corresponding set of coupled DSEs at $\mu=0$(Eq.(4)). However, in practice this is much more complicated than solving Eq.(4). Due to the presence of the $\mu$ in the fourth component
of the momentum variable, the number of arguments of the dressed quark propagator and the independent integration dimension in Eq.(11) is larger than those of Eq.(4) and this fact makes it difficult to find a stationary solutions by means of iteration method. In order to avoid this difficulty, here we shall adopt another more efficient method instead of the iteration method to solve Eq.(11). This is the motivation of our present work. 

Applying the differential operation $\frac{\partial}{\partial(p_4)}$ on  
both sides of Eq.(3), we obtain
\begin{eqnarray}
\frac{\partial G^{-1}(p)}{\partial (p_4)}&=&i\gamma_4+\frac{4}{3}\int \frac{d^4 q}{(2\pi)^4} \frac{\partial}{\partial(p_4)}\left[g^2_{s}D(p-q)\right]\gamma_{\nu}G(q)\gamma_{\nu}. 
\end{eqnarray}

Similarly, applying the differential operation $\frac{\partial}{\partial(p_4)}$ on  
both sides of Eq.(12) successively (n-1)($n\geq 2$) times, we obtain
\begin{eqnarray}
\frac{\partial^n G^{-1}(p)}{\partial (p_4)^n}&=&\frac{4}{3}\int \frac{d^4 q}{(2\pi)^4} \frac{\partial^n}{\partial(p_4)^n}\left[g^2_{s}D(p-q)\right]\gamma_{\nu}G(q)\gamma_{\nu}. 
\end{eqnarray}

Based on Eq.(10) and Eqs.(12-13), we have main results in the present work
\begin{eqnarray}
{\cal G}^{-1}[\mu](p)&=&i\gamma\cdot\tilde{p}A(\tilde{p}^2)+B(\tilde{p}^2)=
G^{-1}(p)+\frac{\partial G^{-1}(p)}{\partial(p_4)}i\mu+\cdots+\frac{1}{n!}\frac{\partial^n G^{-1}(p)}{\partial(p_4)^n}(i\mu)^n+\cdots \nonumber \\
&=&G^{-1}(p)-\mu\gamma_4+\frac{4}{3}\int \frac{d^4 q}{(2\pi)^4}\left\{\frac{\partial}{\partial(p_4)}\left[g^2_{s}D(p-q)\right](i\mu)+\cdots+\right.\nonumber\\
&&\left.\frac{1}{n!}\frac{\partial^n}{\partial(p_4)^n}\left[g^2_{s}D(p-q)\right](i\mu)^n+\cdots\right\}\gamma_{\nu}G(q)\gamma_{\nu}\nonumber\\
&=&G^{-1}(p)-\mu\gamma_4+\frac{4}{3}\int \frac{d^4 q}{(2\pi)^4}\left[g^2_{s}D(\tilde{p}-q)-g^2_{s}D(p-q)\right]\gamma_{\nu}G(q)\gamma_{\nu}\nonumber\\
&=&i\gamma\cdot\tilde{p}+\frac{4}{3}\int \frac{d^4 q}{(2\pi)^4}g^2_{s}D(\tilde{p}-q)\gamma_{\nu}G(q)\gamma_{\nu},
\end{eqnarray}
where we have made use of Eq.(3) in the last step in deriving Eq.(14). If $\mu$ is set to be zero, Eq.(14) reduces to Eq.(3). This is just what one would expect in advance.
At this point a possible place which might lead to misunderstanding should be clarified. In our formulation, the {\it $\mu$-independent} dressed gluon propagator 
$g_s^2D_{\mu\nu}(p)$ is taken as input and the dressed quark propagator at finite
$\mu$ is obtained by solving the rainbow DSE(Eq.(2)). Therefore the explicit presence
of $\mu$ in $g_s^2D({\tilde p}-q)$ does {\it not} imply that the gluon propagator becomes $\mu$-dependent. The $\mu$-dependent ``gluon propagator'' $D({\tilde p}-q)$ is only a quantity resulting from our mathematical trick(Eqs.(12-14)) and should not be interpreted as the actual $\mu$ dependent gluon propagator.

It should be noted that Eq.(14) has two major consequences. First, by means of Eq.(14) the dressed quark propagator at finite chemical potential can be obtained directly from the one at zero chemical potential without the necessity of numerically solving the corresponding integral equations by iteration methods(see Eqs.(15-16) below). This feature facilitates numerical calculations considerably.
Second, Eq.(14) clearly shows that the whole nontrivial $\mu$-dependence of the dressed quark propagator is determined by the $\mu$-dependence of $D({\tilde p}-q)$(the $i\gamma \cdot {\tilde p}$ term only gives the ``trivial'' $\mu$-dependence). This means that, in a complete treatment of the dressed quark propagator at finite chemical potential, the dependence of the dressed gluon propagator on frequency cannot be neglected. Therefore, in all models adopting the instantaneous
approximation(such as the NJL model) the nontrivial $\mu$-dependence of the dressed
quark propagator will be lost(retardation effect are important here).

Multiplying by appropriate gamma matrices and then taking the trace in Eq.(14), we have the following:
\[
[A(\tilde{p}^2)-1]\tilde{p}^2=\frac{8}{3}\int \frac{d^{4}q}{(2\pi)^4}g^2_{s} D(\tilde{p}-q)
\frac{A(q^2)\tilde{p}\cdot q}{q^2A^2(q^2)+B^2(q^2)},
\]
\begin{equation}
B(\tilde{p}^2)=\frac{16}{3}\int \frac{d^{4}q}{(2\pi)^4}g^2_{s} D(\tilde{p}-q)
\frac{B(q^2)}{q^2A^2(q^2)+B^2(q^2)}.
\end{equation}

Substituting $B(p^2)\equiv 0$ into Eq.(15), we have the ``Wigner'' solution for $A'(\tilde{p}^2)$
\begin{equation}
[A'(\tilde{p}^2)-1]\tilde{p}^2=\frac{8}{3}\int \frac{d^{4}q}{(2\pi)^4}g^2_{s} D(\tilde{p}-q)
\frac{\tilde{p}\cdot q}{q^2A'(q^2)},
\end{equation}
The above result simplifies the calculation of $A(\tilde{p}^2)$, $B(\tilde{p}^2)$ and $A'(\tilde{p}^2)$ greatly. Once the form of the model gluon propagator $g^2_sD(p-q)$ is given, one can determine $A(p^2)$, $B(p^2)$ and $A'(p^2)$ by solving the rainbow DSE (Eqs.(4) and (5)). Because the analytic expression for $g^2_sD(p-q)$ is available, one can determine $g^2_sD(\tilde{p}-q)$ by analytic continuation, and thereby obtain $A(\tilde{p}^2)$, $B(\tilde{p}^2)$ and $A'(\tilde{p}^2)$ using Eqs.(15,16) without the necessity of numerically solving the coupled integral equations of
$A(\tilde{p}^2)$ and $B(\tilde{p}^2)$ by means of iteration method.
From the above result one can easily obtain the dressed quark propagator at finite chemical potential from that at zero chemical potential. Here we want to stress that Eqs.(15) and (16) only holds under the ``rainbow'' approximation of DSE and within the radius of convergence of the $\mu$ expansion. In the case of real QCD, it should be noted that both the dressed quark-gluon vertex and the dressed gluon propagator are chemical potential dependent. In this case, Eqs.(15) and (16) would fail. 

Just as was shown in Eqs.(15) and (16), the task of calculating the chemical potential dependence of the dressed quark propagator in the``Nambu-Goldstone'' and the``Wigner'' phase is reduced to the calculation of the three scalar functions $A(p^2)$, $B(p^2)$ and $A'(p^2)$. In order to get the numerical solution of the above three scalar functions, one often use model forms for gluon two-point function as input in Eqs.(4) and (5). As a typical example, we choose the following model gluon propagator;
\begin{equation}
g^2D(q^2)=4\pi^2 d\frac{\chi^2}{q^4+\Delta}~~ with~~ d=\frac{12}{27}.
\end{equation}
The dressed-gluon propagator in Eq.(17) simulates the infrared enhancement and confinement and it leads via the QCD gap equation(Eq.(4)) to an infrared enhancement of the light quark mass function. These modification are intimately related to the confinement and dynamical chiral symmetry breaking[4,5]. The model parameters $\chi$ and $\Delta$ are adjusted to reproduce the weak decay constant in the chiral limit $f_{\pi}=87~MeV$. The forms of $g^2D(q^2)$ have been used in Ref.[13] and it has been shown that with these values a satisfactory description of all low energy chiral observables can be achieved. Substituting Eq.(17) into Eqs.(4) and (5), one can numerically solve the three scalar functions $A(p^2)$, $B(p^2)$ and $A'(p^2)$. The calculated values of the above three scalar functions for model gluon propagator(17) with three sets of different parameters is plotted in Figs.(1,2).

Based on the calculated values of $A(p^2), B(p^2), A'(p^2)$ and Eqs.(15,16), it is not difficult to obtain the corresponding ``Nambu'' and ``Wigner'' solutions at finite chemical potential, i.e., $A(\tilde{p}^2), B(\tilde{p}^2), A'(\tilde{p}^2)$. In order to check whether the numerical solutions $A(\tilde{p}^2), B(\tilde{p}^2)$ obtained from Eq.(15) indeed satisfy the original coupled integral equations(Eq.(11)), one may directly substitute them into Eq.(11) and we find that Eqs.(11) are indeed satisfied. This result can be regarded as a self-consistent check of our numerical calculations in the present work. It is now apparent that our method has the merits that one can easily find the numerical solutions of $A({\tilde p}^2)$ and $B({\tilde p}^2)$ without the necessity of numerically solving Eq. (11) by iteration methods. 
\begin{center}

\epsfig{file=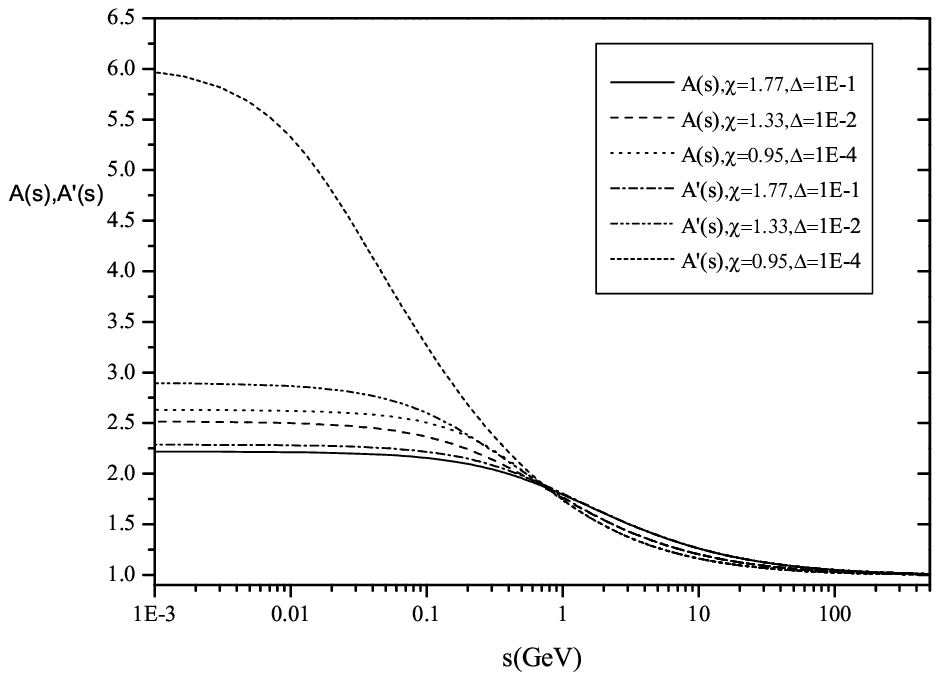, width=9cm}

\vspace{-0.8cm}

Fig.1. $A(s)$ and $A'(s)$ for gluon propagator(17) with three sets of different parameters .

\end{center}

\begin{center}

\epsfig{file=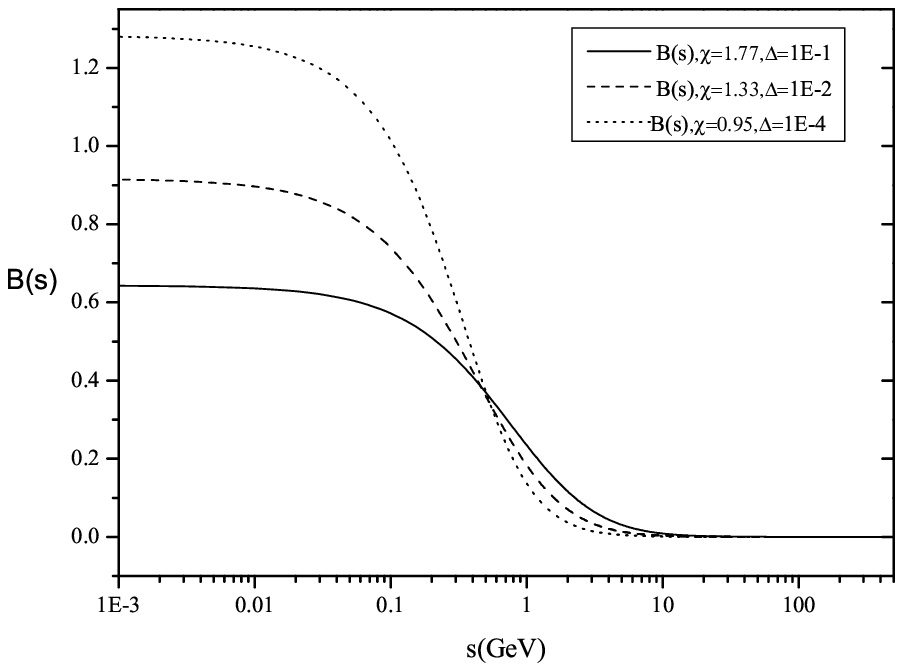, width=9cm}

\vspace{-0.8cm}

Fig.2. $B(s)$ for gluon propagator(17) with three sets of different parameters .

\end{center}

With these two ``phases'' characterized by qualitatively different momentum-dependent quark propagator in the case of non-zero chemical potential, the DSE can be used to explore chiral symmetry restoration and phase transition between the ``Wigner'' and ``Nambu-Goldstone'' phase.

To explore the possibility of a phase transition one must consider the relative stability of the confined and deconfined phase, which is measured by the $\mu$ dependent vacuum pressure difference(or ``bag constant''), which is equivalent to calculating the difference between the tree-level auxiliary-field effective action[14] evaluated with the ``Wigner'' solution characterized by $B(p^2)\equiv 0$, and the ``Nambu-Goldstone'' solution characterised By $B(p^2)\neq 0$[8]:

\begin{eqnarray}
&&{\cal{B}}(\mu)\equiv P[{\cal{G}}^{(NG)}]-P[{\cal{G}}^{(W)}]\nonumber\\
&&=12\int\frac{d p^4}{(2\pi)^4}\left\{ln\left |\frac{\tilde{p}^2A^2(\tilde{p}^2)+B^2(\tilde{p}^2)}
{\tilde{p}^2A'^2(\tilde{p}^2)}\right |+Re\left[\frac{\tilde{p}^2A(\tilde{p}^2)}{\tilde{p}^2A^2(\tilde{p}^2)+B^2(\tilde{p}^2)}-\frac{1}{A'(\tilde{p}^2)}\right]\right\}.
\end{eqnarray}

Substituting $\mu=0$ into Eq.(18), we have the bag constant at zero chemical potential ${\cal{B}}(0)$. By means of numerical studies, the bag constant ${\cal{B}}(0)$ for three different parameter sets of the model gluon propagator(Eq.(17)) are obtained and we list them in Table.I.

\begin{center}
\begin{tabular}{ccc}
\multicolumn{3}{c}{Table. I. The bag constant for the model gluon propagator}\\ \hline\hline
\multicolumn{3}{c}{$g^2D(q^2)$=$4\pi^2d\frac{\chi^2}{q^4+\Delta}$}\\ \hline 
~~~~$\Delta[GeV^4]$~~~~~~~~~~~~&$\chi[GeV]$ ~~~ & ${\cal{B}}^{\frac{1}{4}}(0)[GeV]$ 
\\ \hline
$10^{-1}$ &1.77  &0.122  \\
$10^{-2}$ &1.33  &0.126  \\
$10^{-4}$ &0.95  &0.130  \\ \hline\hline
\end{tabular}
\end{center}

The numerical results for the ratios of
${\cal{B}}(\mu)/{\cal{B}}(0)$ are plotted in Fig.3. The
scale is calculated to be ${\cal{B}}(0)=(0.122~GeV \sim 0.130~GeV)^4$ (see
Table.I), which can be compared with the value $(0.145~GeV)^4$
commonly used in bag-like models of hadron[15].
${\cal{B}}(\mu)>0$ indicates the stability of the confined
phase(Nambu-Goldstone) and hence the phase boundary is specified
by ${\cal{B}}(\mu=\mu_c)=0$. ${\cal{B}}(\mu)$ is positive when the ``Nambu-Goldstone'' phase is
dynamically favored; i.e., has the highest pressure and become
negative when the pressure of ``Wigner'' phase become larger. In Fig.3, we see that ${\cal{B}}(\mu)/{\cal{B}}(0)$ decrease with increasing $\mu$ within a certain range of the small chemical potential(It should be noted that our numerical results is only valid for small values of chemical potential. This is because that our model gluon propagator has no explicit $\mu$ -dependence while the actual gluon propagator should be $\mu$ dependent due to quark vacuum polarizations. As such it may be inadequate at large value of $\mu$). This result is qualitatively the same as that given in Refs.[7,10]. In Refs.[7,10], ${\cal{B}}(\mu)/{\cal{B}}(0)$ decreases with increasing chemical potential up to $\mu_c$.

\begin{center}

\epsfig{file=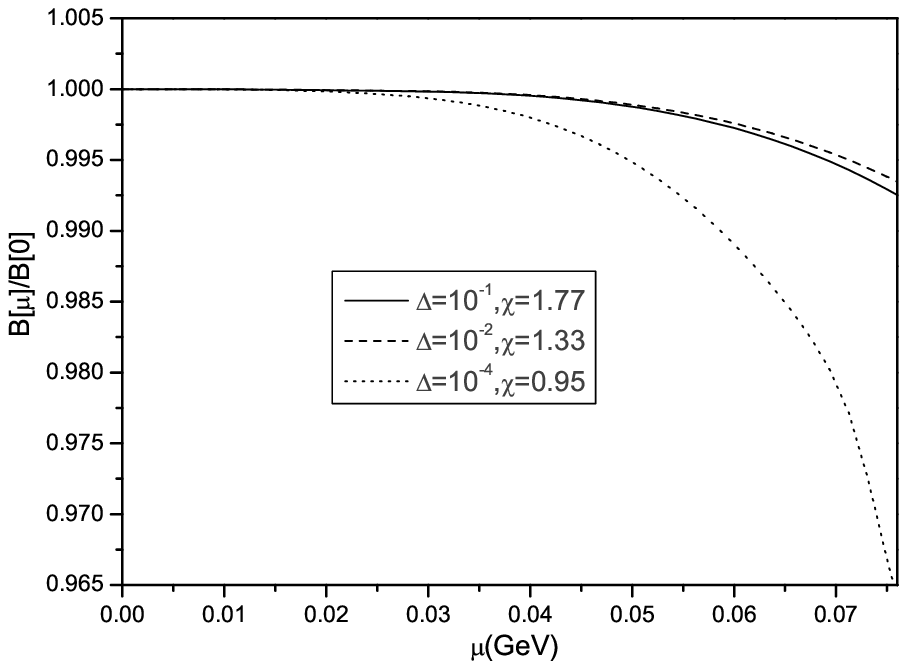, width=8cm}

\vspace{-0.8cm}

Fig.3. Ratio $B(\mu)/B(0)$ as a function of $\mu$.

\end{center}

Now let us turn to the study of the measure of the dynamical
chiral symmetry breaking in the case of non-zero chemical potential. In
order to get a reasonable result for the mixed quark-gluon
condensate and vacuum susceptibilities in an effective quark-quark
interaction model, the authors in Refs.[16,17] defined the
``effective'' two-quark condensate as the difference between the
``exact'' quark propagator(quark propagator in the
``Nambu-Goldstone'' phase, in which chiral symmetry is dynamically
broken and the dressed quarks are confined) and the
``perturbative'' quark propagator(quark propagator in the "Wigner"
phase, in which chiral symmetry is not dynamically broken and the
dressed quarks are not confined). It can be written as (in the chiral
limit and at zero chemical potential):
\begin{eqnarray}
&&\langle\tilde{0}|\bar{q}q|\tilde{0}\rangle_{\mu=0}\equiv -tr_{DC}
\left\{{\cal{G}}^{(NG)}[\mu=0]-{\cal{G}}^{(W)}[\mu=0]\right\}.
\end{eqnarray}
It should be noted that Eq.(19) is only valid in an effective
quark-quark interaction model(more details can be found in
Ref.[18]).

Here we extend the above concept to get a measure of dynamical
chiral symmetry breaking in the case of finite $\mu$ and
obtain the ``effective''  two-quark condensate with the finite $\mu$ as:
\begin{eqnarray}
\langle\tilde{0}|\bar{q}q|\tilde{0}\rangle_{\mu}\equiv -tr_{DC}
\left\{Re{\cal{G}}^{(NG)}[\mu]-Re{\cal{G}}^{(W)}[\mu]\right\}
=-12\int\frac{d p^4}{(2\pi)^4} Re\left[\frac{B(\tilde{p}^2)}{\tilde{p}^2A^2(\tilde{p}^2)+B^2(\tilde{p}^2)}\right].
\end{eqnarray}
Substituting $\mu=0$ into Eq.(20), we have the usual ``effective'' two-quark condensate in the chiral limit. The calculated ratio
$\langle\tilde{0}|\bar{q}q|\tilde{0}\rangle_{\mu}/\langle\tilde{0}|\bar{q}q|\tilde{0}\rangle_{\mu=0}$ is plotted in Figs.(4,5). From Figs.(4,5) we can see that the behavior of $\langle\tilde{0}|\bar{q}q|\tilde{0}\rangle_{\mu}/\langle\tilde{0}|\bar{q}q|\tilde{0}\rangle_{\mu=0}$ depends strongly on the choice of parameters in the model gluon propagator. For the first set of parameters $\langle\tilde{0}|\bar{q}q|\tilde{0}\rangle_{\mu}/\langle\tilde{0}|\bar{q}q|\tilde{0}\rangle_{\mu=0}$ decreases monotonically as $\mu$ increases, while for the second and third sets of parameters $\langle\tilde{0}|\bar{q}q|\tilde{0}\rangle_{\mu}/\langle\tilde{0}|\bar{q}q|\tilde{0}\rangle_{\mu=0}$ first increases with increasing $\mu$, and after reaching 
a crest, decreases with increasing $\mu$. This results is quite different from that given by Refs.[7,10,19-20]. In Refs.[7,10], one claims that $\langle\tilde{0}|\bar{q}q|\tilde{0}\rangle_{\mu}/\langle\tilde{0}|\bar{q}q|\tilde{0}\rangle_{\mu=0}$ increases with increasing $\mu$ up to $\mu_c$ as a consequence of the necessary momentum-dependence of the dressed-quark self-energy.

\begin{center}

\epsfig{file=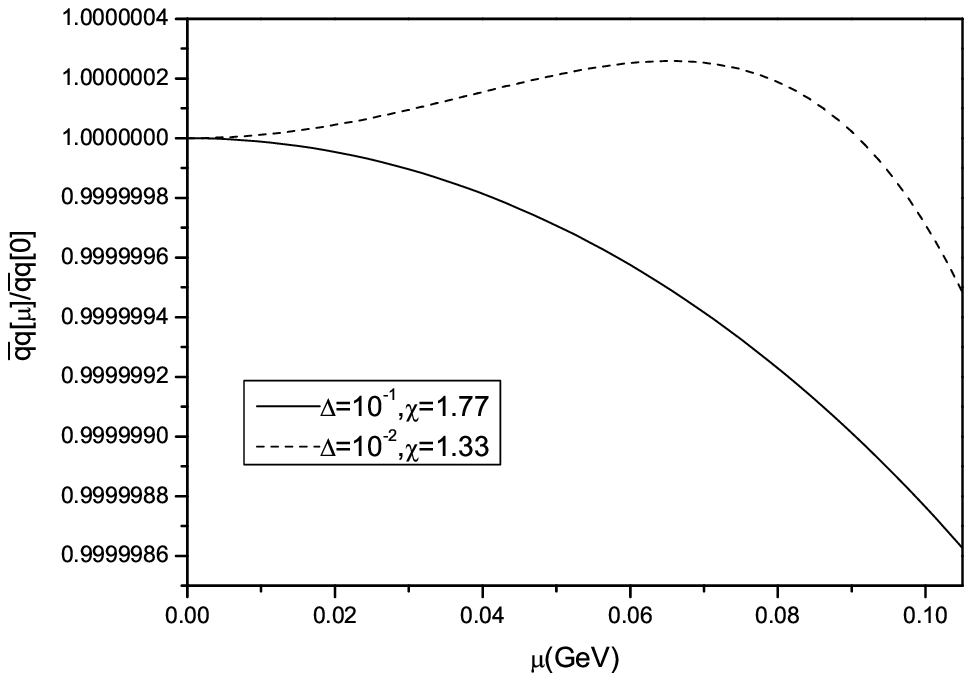, width=9cm}

\vspace{-0.7cm}

Fig.4. The ratio
$\langle\tilde{0}|\bar{q}q|\tilde{0}
\rangle_{\mu}/\langle\tilde{0}|\bar{q}q|\tilde{0}\rangle_{\mu=0}$ as a function of $\mu$.

\end{center}

\begin{center}

\epsfig{file=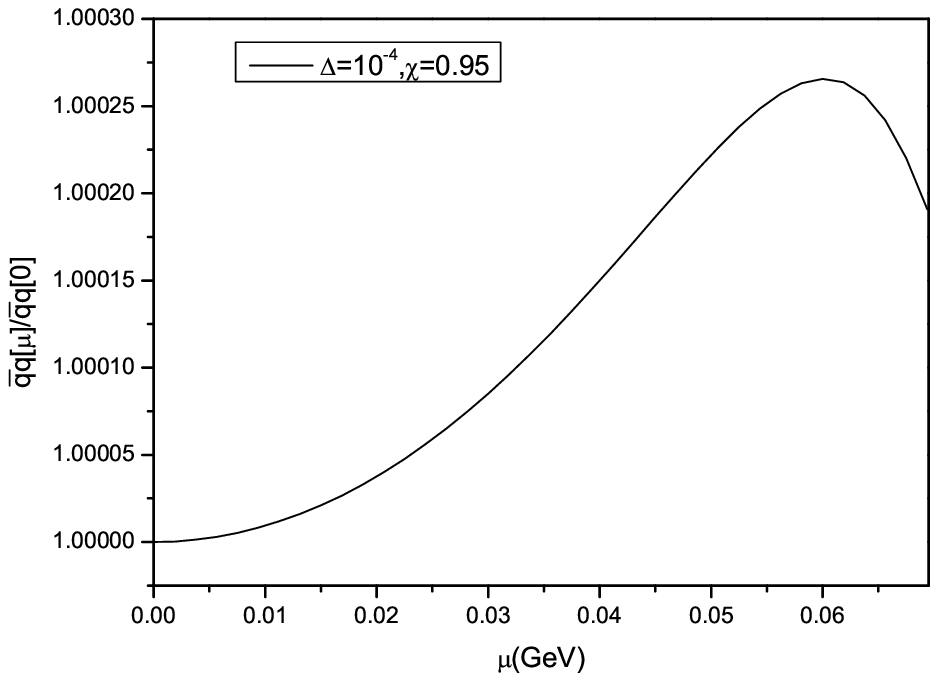, width=9cm}

\vspace{-0.7cm}

Fig.5. The ratio
$\langle\tilde{0}|\bar{q}q|\tilde{0}
\rangle_{\mu}/\langle\tilde{0}|\bar{q}q|\tilde{0}\rangle_{\mu=0}$ as a function of $\mu$.

\end{center}

It should be noted that this strong dependence on model parameters is not unexpected because the whole nontrivial $\mu$-dependence of dressed quark propagator ${\cal G}^{-1}[\mu](p)$ is determined by the $\mu$-dependence of $D({\tilde p}-q)$(Eq.(14)). In practice this can be used as guide for choice of model parameters. Based on the Hellmann-Feynman theorem and Gell-Mann, Oakes, Renner relations(PCAC), one finds:
\begin{equation}
\frac{\langle\tilde{0}|\bar{q}q|\tilde{0}
\rangle_{\rho}}{\langle\tilde{0}|\bar{q}q|\tilde{0}\rangle_{\rho=0}}=1-\frac{\rho\sigma_N}{f^2_{\pi}m^2_{\pi}}+O(\rho^2).
\end{equation}
The leading-order linear baryon density($\rho$) dependence is determined by the nucleon sigma term in a model-independent way(for details see Ref.[21]),
 i.e., $\sigma_N=\langle N|m_q(uu+dd)|N \rangle$. If we regard Eq.(21) as a criterion for studying the chiral order parameter at finite density, then one will find the second and third sets of parameters of the model gluon propagator(17) is not suitable.

To summarize: in the present work, we provide a general recipe to
calculate the chemical potential dependence of the dressed quark
propagator under the rainbow approximation to the DS equation. This approach has the advantage that the dressed quark propagator at finite chemical potential can be obtained directly from the one at zero chemical potential without the necessity of
numerically solving the corresponding coupled integral equations by iteration methods.
This feature facilitates numerical calculations considerably. From this the "effective" quark
condensate at finite chemical potential and the phase transition between the ``Wigner'' and ``Nambu-Goldstone'' phase is analyzed.  It is found that the behavior of the "effective" quark condensate $\langle\tilde{0}|\bar{q}q|\tilde{0}\rangle_{\mu}/\langle\tilde{0}|\bar{q}q|\tilde{0}\rangle_{\mu=0}$ depends strongly on the choice of parameters of the model gluon propagator. This can serve as a guide for choosing model parameters in further study. 
Finally we want to stress that the approach adopted in this letter is general in principle and can be applied to the study of the color superconductivity in the framework of rainbow DS approach[22].

\vspace*{0.8 cm}

\noindent{\large \bf Acknowledgments}

This work was supported in part
by the National Natural Science Foundation of China(under Grant
Nos 10425521, 10175033, 10135030) and the Research Fund for the Doctoral
Program of Higher Education(under Grant Nos 20030284009 and 20040001010).

\vspace*{0.8 cm}

\noindent{\large \bf References}

\begin{description}
\item{[1]} Z. Fodor and S. D. Katz, JHEP {\bf 0404}, 050 (2004); JHEP {\bf 0203}, 014 (2002).
\item{[2]} C. R. Allton et al., Phys. Rev. {\bf D 66}, 074507 (2002).
\item{[3]} P. de Forcrand and O. Philipsen, Nucl. Phys. {\bf B 642}, 290 (2002); M. D'Elia and M. P. Lombardo, Phys. Rev. {\bf D 67}, 014505 (2003)
\item{[4]} C. D. Roberts and A. G. Williams, Prog. Part. Nucl. Phys. {\bf 33}, 477 (1994), and references therein.
\item{[5]} P. C. Tandy, Prog. Part. Nucl. Phys. {\bf 39}, 117 (1997);  R. T. Cahill and S. M. Gunner, Fiz. {\bf B7}, 17 (1998), and references therein.
\item{[6]} R. Alkofer, L. von Smekal, Phys. Rep. {\bf 353}, 281 (2001), and references therein.
\item{[7]} C. D. Roberts and S. M. Schmidt, Prog. Part. Nucl. Phys. {\bf 45S1},
1 (2000), and references therein.
\item{[8]} D. Blaschke, C. D. Roberts, and S. Schmidt, Phys. Lett. {\bf B425}, 232 (1998).
\item{[9]} P. Maris, C. D. Roberts, and S. Schmidt, Phys. Rev. {\bf C57}, R2821 (1998).
\item{[10]} A. Bender, W. Detmold, and A. W. Thomas, Phys. Lett. {\bf B516}, 54 (2001).
\item{[11]} Hong-shi Zong, Feng-yao Hou, Xiang-song Chen, and Yu-xin Liu, Chin. Phys. Lett. {\bf 21}, 1232 (2004); Hong-shi Zong, Feng-yao Hou, Wei-min Sun, and Xiao-hua Wu, Commun. Theor. Phys. (Beijing, China) {\bf 42}, 581 (2004).
\item{[12]} Hong-shi Zong, Chang Lei, Feng-yao Hou, Wei-min Sun and Yu-xin Liu, Phys. Rev. {\bf C71}, 015205 (2005).
\item{[13]} T. Meissner,  Phys. Lett. {\bf B405}, 8 (1997).
\item{[14]} R. W. Haymaker, Riv. Nuovo Cim. {\bf 14}, series 3, No. 8 (1991).
\item{[15]} R. T. Cahill, Aust. J. Phys. {\bf 42}, 171 (1989).
\item{[16]} Hong-shi Zong, Jia-lun Ping, Hong-ting Yang, Xiao-fu L\"{u} and Fan Wang, Phys. Rev. {\bf D67}, 074004 (2003).
\item{[17]} Hong-shi Zong, Shi Qi, Wei Chen, Wei-min Sun, and En-guang Zhao, Phys. Lett. {\bf B576}, 289 (2003).
\item{[18]} P. Maris, C. D. Roberts, P. C. Tandy, Phys. Lett. {\bf B420}, 267 (1998).
\item{[19]} Y. Taniguchi and Y. Yoshida, Phys. Rev. {\bf D55}, 2283 (1997).
\item{[20]} O. Kiriyama, M. Maruyama, and F. Takagi, Phys. Rev. {\bf D62}, 105008 (2000).
\item{[21]} W. Weise, Nuclear Physics. {\bf A610}, 35c (1996), and references therein.
\item{[22]} Hong-shi Zong, et.al., in preparation.
\end{description}

\end{document}